# Parallelism of measures in non-Markovian environments

Abolfathiyar Masoud[1,*], Entekhabi Sina[2]

[1]*Department of Physics, Faculty of Sciences, Azerbaijan University of Shahid Madani, Tabriz, Iran*

[2]*Department of Computer Sciences, Faculty of Engineeingr, Urmia University, Urmia, Iran*

[*]*Email: mphysic1365@gmail.com*

## 1. Abstract

There are various measures to compute correlation in various conditions, but none of them can be used publicly, because each of them has abnormality in their calculations or behaviors. These abnormalities do not depend on genus of environments. Therefore, in this work, we will show correlation diversity of evolution that is computed by each measure for two oscillators with their non-Markovian stochastic reservoir. We will see that while the situation is the same for all of them, they will have symmetrical plots. In more expressive words, process of the measures has a kind of parallelism when angular momentum and time are increasing. In the other hand, increase of quantum correlation under the non-Markovian drive is observable in all of studied measures. Also, in all of them (quantum discord's measures and entanglement's measures) recently occurrences are done.

## 2. Introduction

We know that one of the important properties of the quantum systems is when they are correlated, for in these form, classical relations cannot justify their behaviors. So, you have to use some measures such as entanglement [1], quantum discord [2-10], the negativity [11,12] and else, to exhibit quantum correlation.
In quantum systems a factor that is called "noise" can cause some abnormal behaviors of the correlation [28,29], that this subject is perused as a scrutiny of entanglement dynamics for two-qubit systems [28-32], harmonic oscillators [33,34], and even recently in non-Markovian environments [25].

Often dissipative environments (such as noisy-type) are hard to treat analytically and are not feasible by standard techniques [41]. Therefore, numeric or impose approximations should be used to obtain flexible results [25] (the non-Morkovian approximation is used here). If the bath was considered Markovian (memoryless), due to totally uncorrelated interactions of the system's quantum state with the bath, the memory is wiped away, and the destruction of entanglement can be rather swift [25]. Therefore, it can be pointed to the prolonging of entanglement and its rebirth as the non-Markovian's bath effects [25].

But, to compute for the behaviors of the oscillators in the presence of non-Markovian drive, before anything, the noise-averaged density matrix must be obtained (that is obtained in [25] and here is Eq.(4)). For two-oscillator form that coupling each of them with a noise source separately is the only independence reason of oscillators, due to each noisy source, some amount of energy ($\grave{o}$) is added to them. So, according to what mentioned above, the density matrix will be a function of the added energy ($\grave{o}$), and all memory effects enter only through the energy. This means that non-Markovian effects are entered by the added energy, and we study the system's behavior with respect to ($\grave{o}$).

Although we can show the rate of correlation by using each of its measures, but there are some problems for them, such as having sudden death (or revive), or obtaining by approximate method, or showing abnormal behaviors, and else. In the following, first it will be said that under what conditions the density matrix is obtained, so we will present relationships of measures (the correlation measures and the entanglement measures) very briefly. Finally we will calculate all of them for the above mentioned system.

## 3. The density matrix

Let us consider a system that is composed of two oscillators each interacts with non-Markovian resource [25], so their Hamiltonian can be written as follows:

$$\widehat{H} = \omega\left(a^\dagger a + \frac{1}{2}\right) + \frac{1}{\sqrt{2}}[\xi(t)a^\dagger + h.c.] + \omega\left(b^\dagger b + \frac{1}{2}\right) + \frac{1}{\sqrt{2}}[\eta(t)b^\dagger + h.c.] \quad (1)$$

where $\xi(t)$ and $\eta(t)$ define our noise. According to (1) time evolution of the density matrix can be obtained as follows:

$$\hat{\rho}(t) = e^{\mathcal{L}_1(t)} \otimes e^{\mathcal{L}_2(t)} \hat{\rho}(0) \quad (2)$$

where $\mathcal{L}_i$ are "Liouvillian" operators [26] and $\hat{\rho}(0)$ is the initial density matrix that it is a maximally entangled X-state because X-type density matrices calculations are explicit, then the initial state is selected so that the final density matrix be X-type. Also, the main objective is showing procedure of effects of non-Markovian environments on the correlation better than usual. Accordingly, the initial state is maximally entangled to realize our aim.

$$\hat{\rho}(0) = \frac{1}{2}(|01\rangle + |10\rangle) \otimes (\langle 01| + \langle 10|) \quad (3)$$

Finally, the density matrix for two independent harmonic oscillators each coupled to its own stochastic noise reservoir is obtained as follows (without normalizing) [25]:

$$\hat{\rho}(t) = \begin{pmatrix} \dfrac{\dot{o}(t)}{(1+\dot{o}(t))^3} & 0 & 0 & 0 \\ 0 & \dfrac{\frac{1}{2}+\dot{o}(t)^2}{(1+\dot{o}(t))^4} & \dfrac{\frac{1}{2}}{(1+\dot{o}(t))^4} & 0 \\ 0 & \dfrac{\frac{1}{2}}{(1+\dot{o}(t))^4} & \dfrac{\frac{1}{2}+\dot{o}(t)^2}{(1+\dot{o}(t))^4} & 0 \\ 0 & 0 & 0 & \dfrac{\dot{o}(t)(1+\dot{o}(t)^2)}{(1+\dot{o}(t))^5} \end{pmatrix} \qquad (4)$$

It should be noted that the main variable is $\epsilon$ as the added energy to the system occurs due to noise (the below figure shows time evolution of $\epsilon$ in three noisy environments, therefore, it can be attended to relation between this evolution and measures evolutions), and it has different behaviors in different conditions, for example in *white noise, partially colored noise* and *colored noise* environments.

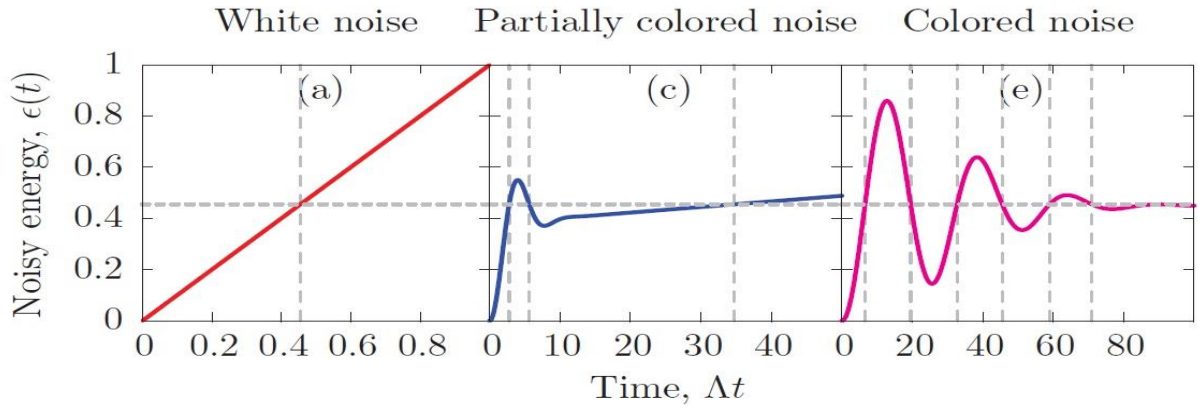

**Figure 1**: As is obvious, the mentioned fluctuations for added energy (which accordingly the measures experience change) occur in the proximity of $\dot{o}(t) = 0.455\omega$. As so these fluctuations have been figured vividly in [25] for entanglement.

BE CAREFUL; *in all calculations, the energy computation is fulfilled by using Gaussian Rule (3-point method) approximation*.

### 3. The correlation measures

*Quantum discord* (QD) as a correlation measure [35], in the Markovian (or non-Markovian) environments is stronger and comprehensive than entanglement (E) [36]. After obtaining graphs of this measure for our system in three terms (white noise, partially colored noise, and colored noise that are plotted in FIG4, FIG5, and FIG6 respectively), we shall see presence of QD even when E is dead. Even if common environments are used, the significant differences between behaviors of QD and E could be observed again [37].

*Geometric measure* that makes calculations of QD to become easier [39], and although it is a measure with one certain way relativity, but there are some critical points that their mathematical reasons are clear, though their physical reasons have not been found [38]. This measure, also is plotted in those above mentioned terms (FIG 7, FIG 8, and FIG 9)

## 4. Entanglement monotone

*Negativity (N)* is a correlation measure, and it is retrieved as an entanglement monotone. It is discussed for quantum channels, too [40], and is plotted for our terms (FIG 10, FIG 11, and FIG 12). Thereinafter, *logarithmic negativity ($E_N$)* is an entanglement monotone like the negativity (and entanglement measure) [11], then properties of entanglement monotone and additional have been explored [13-24]. $E_N$, it has the following properties [27]:

- can be zero even if the state is entangled (if the state is PPT≡Positive Partial Transpose preserving, entangled)
- does not reduce to the entropy of entanglement on pure states like most other entanglement measures
- is additive on tensor products: $E_N(\rho \otimes \sigma) = E_N(\rho).E_N(\sigma)$
- is not asymptotically continuous. That means that for a sequence of bipartite Hilbert spaces $H_1, H_2,...$ (typically with increasing dimension) we can have a sequence of quantum states $\rho_1, \rho_2,...$ which converges to $\rho^{\otimes n_1}, \rho^{\otimes n_2},...$ (typically with increasing $n_i$) in the trace distance, but the sequence $E_N(\rho_1)/n_1, E_N(\rho_2)/n_2,...$ does not converge to $E_N(\rho)$.

- is an upper bound to the distillable entanglement.

$E_N$ is defined by [11,28]:

$$E_N(\rho) = LN(\rho) = log_2 \|\rho^{\Gamma_A}\|_1 \quad (5)$$

where $\|A\|_1 = tr|A| = tr\sqrt{A^\dagger A}$ and $\rho^{\Gamma_A}$ is the partial transpose of $\rho$ with respect to party A (FIG 1, FIG 2, and FIG 3).

## 4. Summary and conclusion

In this work, a comparison has been done between behaviors of the correlation measures and also entanglement measures. We have used the system that includes two independent oscillators under non-Markovian drive, and it was associated with an energy that has been added to the system due to noise [25]. Mentioned energy is our main variable because as it was seen, behavioral changes of measures depended on the rate of it, directly. However, the following results can be noted:

*a)* Process of the measures is the same for increase of angular momentum and time. G has improved QD as we saw proceeding mentioned between N and $E_N$, respectively. It should be noted that, mechanisms of G action is exactly opposite of the $E_N$ mechanisms.

*b)* N and $E_N$ are the entanglement measures, but it can be seen that these values are not zero even when E is zero and vice versa.

*c)* In all of them (the correlation measures and the entanglement measures), they decay to zero when the frequency and time increases. At low frequencies (and initial time), especially when the energy of the noise has approximately large fluctuations, they have most of the changes.

*d)* Changes of them (all of the measures) are different from an environment to another, or in other words, each environment has more noise, it will lead to further changes in low frequencies (or initial times).

*e)* Noise increase (non-Markovian one) causes the revive of correlation or its increase, which is observable in figures relating to each measures. As such, in environment with more noise the amount of correlation has been more than environment with less noise.

*f)* In white noise which energy increase is linear less correlation is seen while in other environments due to noise increase, the energy experience fluctuation correlation increases and in some occasions we encounter the revival of correlation. It can be explained through $\omega$. By increasing $\omega$ the fluctuated mood of energy will also face more fluctuations. As fluctuations number increase, the correlation amount will increase too. Accordingly, in environment where the noise enters the system in fluctuated mood the correlation would lose its existence with fewer and has enjoy much more amount than environments where noise enters linear mood into system.

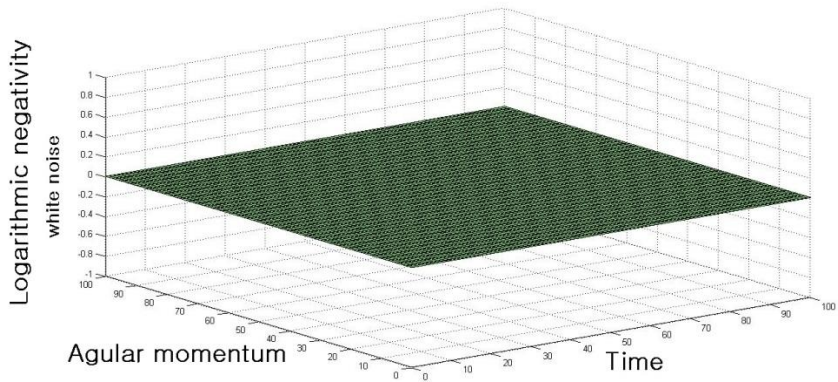

Figure 2: Changes of $E_N$ white noise environment with non-Markovian conditions ($\Lambda\tau = 0$)

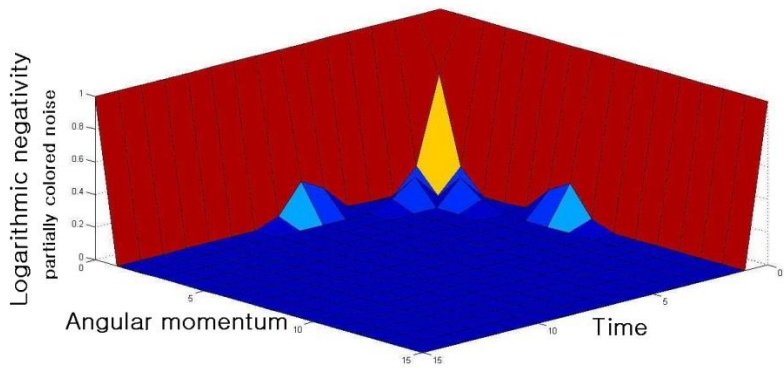

Figure 3: Changes of $E_N$ partially colored noise environment with non-Markovian conditions ($\omega\tau = 3.5, \frac{\omega}{\Lambda} = 0.875$)

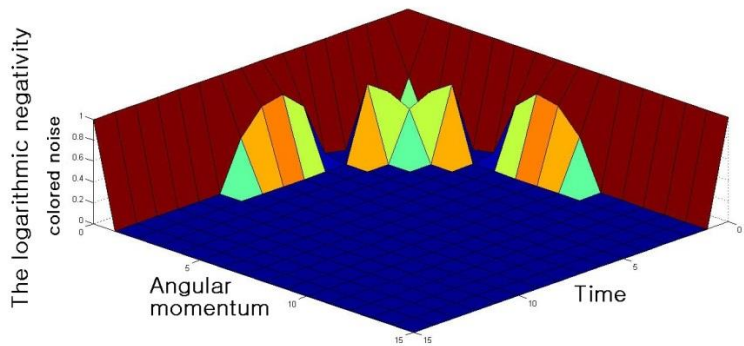

Figure 4: Changes of $E_N$ colored noise environment with non-Markovian conditions ($\omega\tau = 7.5, \frac{\omega}{\Lambda} = 0.25$)

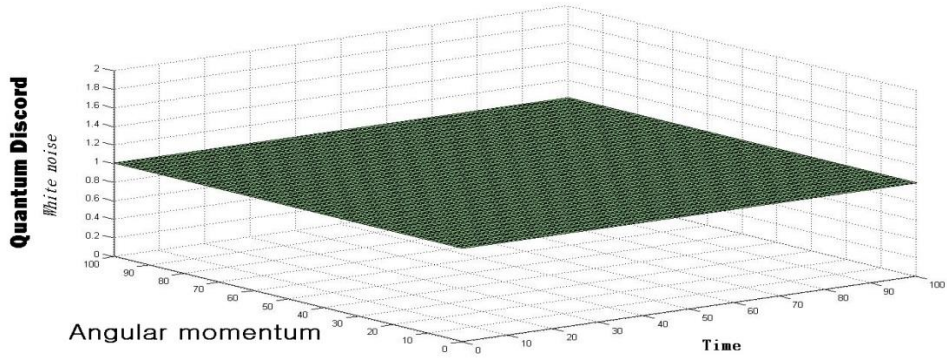

**Figure 5: Changes of QD in partially colored noise environment with non-Markovian conditions ($\Lambda\tau = 0$)**

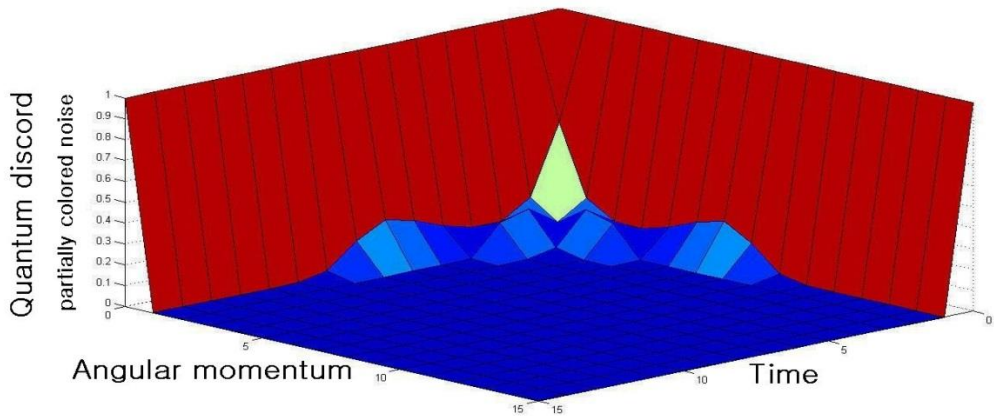

**Figure 6: Changes of QD in partially colored noise environment with non-Markovian conditions ($\omega\tau = 3.5, \frac{\omega}{\Lambda} = 0.875$)**

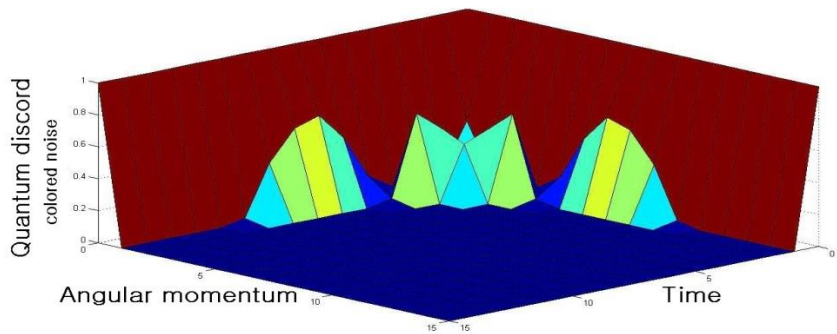

**Figure 7: Changes of QD in partially colored noise environment with non-Markovian conditions ($\omega\tau = 7.5, \frac{\omega}{\Lambda} = 0.25$)**

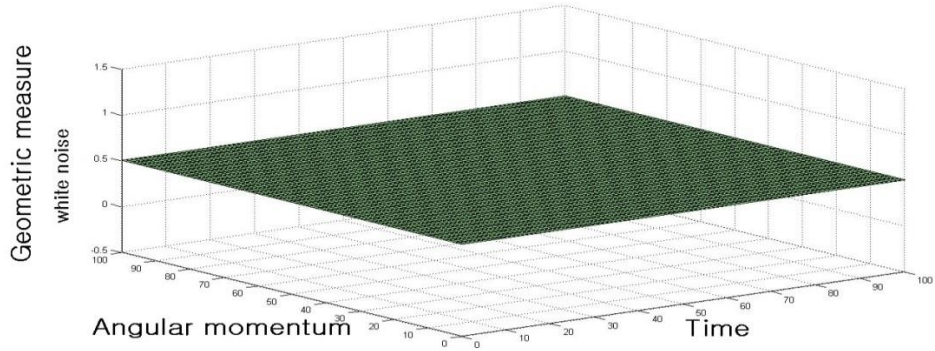

**Figure 8: Changes of $G$ in partially colored noise environment with non-Markovian conditions ($\Lambda\tau = 0$)**

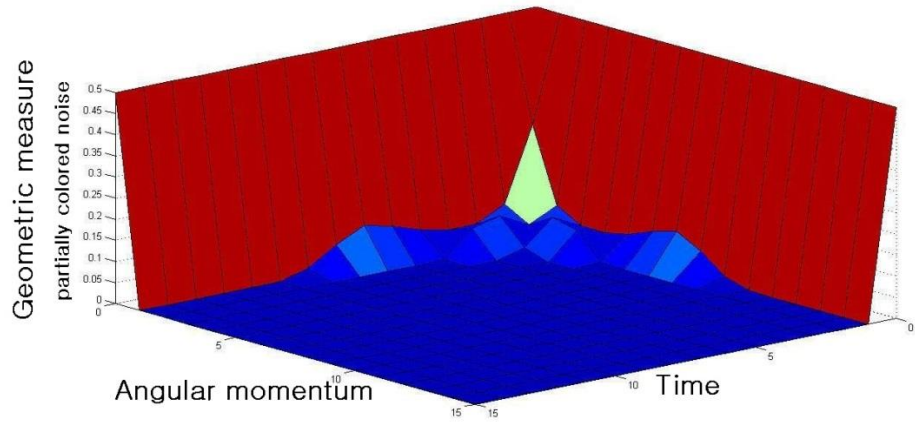

**Figure 9: Changes of $G$ in partially colored noise environment with non-Markovian conditions ($\omega\tau = 3.5, \frac{\omega}{\Lambda} = 0.875$)**

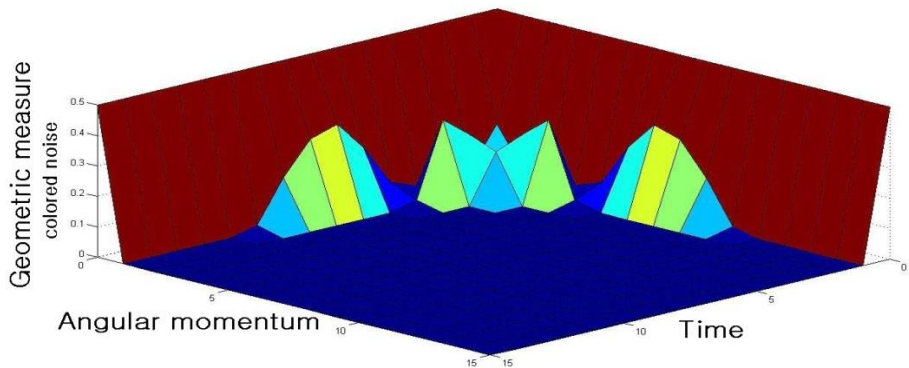

**Figure 10: Changes of G in partially colored noise environment with non-Markovian conditions ($\omega\tau = 7.5, \frac{\omega}{\Lambda} = 0.25$)**

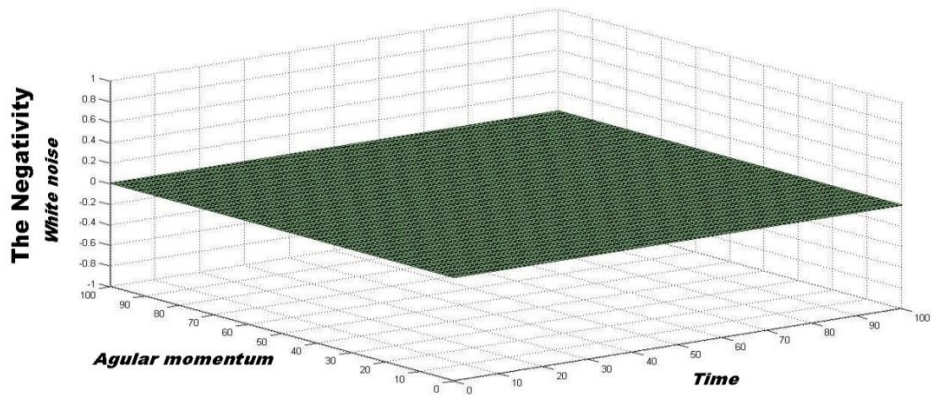

**Figure 11:** Changes of N in partially colored noise environment with non-Markovian conditions ($\Lambda\tau = 0$)

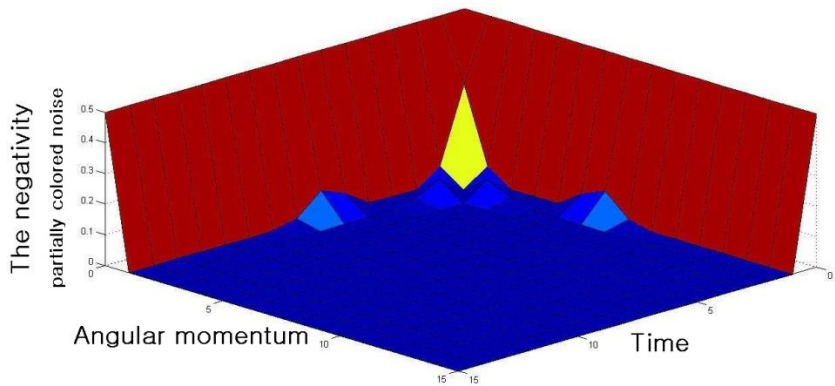

**Figure 12:** Changes of N in partially colored noise environment with non-Markovian conditions ($\omega\tau = 3.5, \frac{\omega}{\Lambda} = 0.875$)

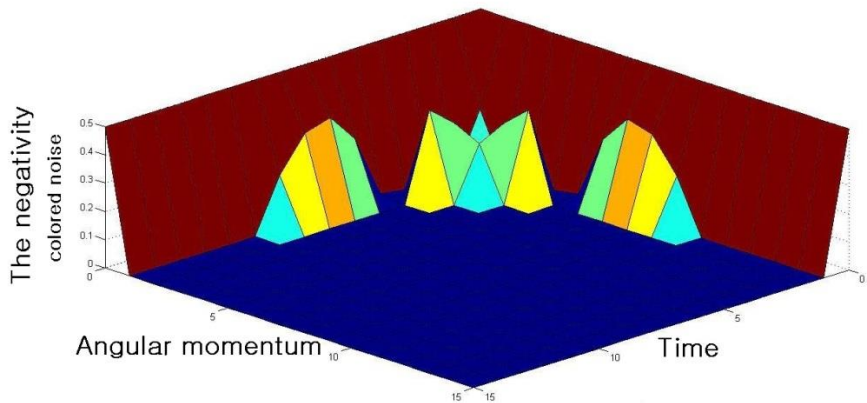

**Figure 13:** Changes of N in partially colored noise environment with non-Markovian conditions ($\omega\tau = 7.5, \frac{\omega}{\Lambda} = 0.25$)